\documentclass[%
 reprint,
groupedaddress,
showpacs,
 amsmath,amssymb,
 aps,
 prd,
 longbibliography,
 lengthcheck,%
]{revtex4-1}

\usepackage{graphicx}
\usepackage{dcolumn}
\usepackage{bm}
\usepackage{float}

\begin{document}

\title{Solutions of Mathisson-Papapetrou equations \\
for  highly relativistic spinning particles}

\author{Roman Plyatsko, Mykola Fenyk, and Oleksandr Stefanyshyn}

\affiliation{Pidstryhach Institute for Applied Problems in
Mechanics and Mathematics\\ Ukrainian National Academy of Sciences, 3-b Naukova Street,\\
Lviv, 79060, Ukraine}


\begin{abstract}
Different types of essentially nongeodesic motions of highly relativistic spinning particles in Schwarzschild's and Kerr's background which follows from the Mathisson-Papapetrou (MP) equations are considered. It is shown that dependently on the correlation of signs of the spin and the particle's orbital velocity the spin-gravity coupling acts as a significant repulsive or attractive force. Numerical estimates for electrons, protons, and neutrinos in the gravitational field of black holes are presented. The correspondence between the general relativistic Dirac equation and  MP equations is discussed. It is stressed that for the highly relativistic motions the adequate supplementary condition for the MP equations is the Mathisson-Pirani condition. In the following it is important to study the possible role of the highly relativistic spin-gravity coupling in astrophysics, cosmology, and high
energy physics.

\end{abstract}


\maketitle

\section{ Introduction}

After obtaining the general relativistic equations of motion for a test body with inner rotation/nonquantum spinning particle in [1, 2],  investigations of their solutions for some backgrounds were started in [3--7].  Schwarzschild's metric was taken into account in [3, 5, 6],  Lense-Thirring's and Melvin's metric were under consideration  in [4] and [7] correspondingly. The concrete physical effects of  spin-gravity coupling on rotating body trajectories have been studied. For example,  it was shown that the correction in perihelion motion of the planet Mercury because of its spin is negligible [3]. Similar results were obtained for other cases of  spinning particle motions in the gravitational field which were considered at that time. Therefore, it is naturally that the corresponding statement was formulated in the known book [8] from 1973. However, the same year is dated paper [9] where another supposition one can read: "The simple act of endowing a black hole with angular momentum has led to an unexpected richness of possible physical phenomena. It seems appropriate to ask whether endowing the test body with intrinsic spin might not also lead to surprises."
This paper,  together with [10], where the spin-spin and spin-orbit gravitational interactions were considered, gave the impulse for realizing  our program of more detailed investigations of physical effects following from the MP equations. We began from searching possible new analytical solutions of the exact Mathisson-Papapetrou equations in Schwarzschild's background without the supposition that they must be close
to the corresponding solutions of the geodesic equations. That is, in our calculations we did not use the approach of small corrections because of the spin to the known geodesic solutions.

It is known that the exact MP equations have a simple analytical solution for the radial motion with any oriented spin in Schwarzschild's background: this solution describe the particle's world line which coincides with the corresponding geodesic world line. Other solutions of the MP equations in Schwarzschild's background, which describe the circular orbits of a particle with spin orthogonal to the equatorial plane of its motion, is known as well. These solutions differ from the corresponding geodesic solutions because of the terms which describe the gravitational spin-orbit interaction. At the same time, it is not difficult to check that the MP equations do not admit solutions in Schwarzschild's equatorial plane with the  spin non-orthogonal to this plane. However, paper [10] inspired the question: do the MP equations admit any solution which describe circular orbits of a spinning particle beyond the equatorial plane of
 Schwarzschild's background? Indeed, it was shown in [10] that according to the MP equations the spin-spin interaction cannot maintain a spinning test particle at rest on the axis of rotation of Kerr's mass. Nevertheless, this fact does not exclude a priori an effect of particle hovering above Kerr's or Schwarzschild's mass due to the spin-orbit interaction, i.e., for some dynamical, nor statical, cases. Then it was shown
that the MP equations admit the corresponding solutions both in the Schwarzschild  [11, 12] and Kerr backgrounds [13] which describe just the circular non-equatorial orbits. It is important that for realizing these orbits a spinning particle must posses the highly relativistic velocity relative to a source of the gravitational field. Late those and other results of studying the highly relativistic spinning particle motions were summarized in book [14]. Further development of the corresponding direction of investigations for the last 15 years is presented in [15--23].

The important point in the description of spinning particle motions by the MP equations is the choice of an appropriate supplementary condition. From the first steps in our investigations we chose the Mathisson-Pirani supplementary condition  [1, 24] as a basic one.
In this choosing we took into account the results of papers  [25, 26]: if we need to describe, in the proper sense, just the inner rotation of the body, it is necessary to use the Mathisson-Pirani condition. Nevertheless, in many cases, when the body/particle velocity is not very close to the velocity of light, this condition can be substitute by the Tulczyjew-Dixon condition with high accuracy (more on this subject we write below).

Here we present our results of the theoretical investigations of the highly relativistic spin-gravity coupling by  analysis of the corresponding solutions of the MP equations in Schwarzschild's and Kerr's background. Naturally, from a practical point of view the situation with a macroscopic test particle moving relative to a massive body with the velocity close to the velocity of light is not realistic. However, the highly relativistic velocities are usual in astrophysics for elementary particles. In this connection we stress the important fact: in many papers [27--35] it was shown that in a certain sense the MP equations follow from the quantum general relativistic Dirac equation [36--38] as a classical approximation.

\section{Different representations \\ of Mathisson-Papapetrou equations}

Because the traditional form of the MP equations are presented above in other papers, in this section we write only their non-traditional forms which are convenient for further analysis of  physical meaning of the corresponding solutions.

\subsection{ MP equations through spin 3-vector}

For description of the particle spin by the MP equations, in many papers both the tensor of spin $S^{\lambda\mu}$ and the 4-vector of spin $s_\lambda$ are used, where by definition
\begin{equation}
\label{plyatsko:1}
s_\lambda=\frac 12 \varepsilon_{\lambda\mu\nu\sigma} \sqrt{-g} u^\mu S^{\nu\sigma}, \quad
S^{\lambda\mu}=\varepsilon^{\lambda\mu\nu\sigma} u_\nu s_\sigma,
\end{equation}
where $\varepsilon_{\lambda\mu\nu\sigma}$  and $g$  are the Levi-Civita symbol and the determinant of the metric tensor correspondingly (here, and in the following, Greek indices run 1, 2, 3, 4 and Latin indices 1, 2, 3; the signature of the metric $- , - , - , +$ is chosen; units $c=G=1$ are used) .
That is, instead of the six components $S^{\lambda\mu}$ one can deal with the four components $s_\lambda$. In the term of $s_\lambda$ the Mathisson-Pirani condition is
\begin{equation}
\label{plyatsko:2}
s_\lambda u^\lambda =0,
\end{equation}
and the spin part of the MP equations takes the form
\begin{equation}
\label{plyatsko:3}
\frac {Ds^\lambda}{ds}=s_\mu \frac {Du^\mu }{ds} u^\lambda,
\end{equation}
i.e., according to the MP equations at the Mathisson-Pirani the 4-vector of spin is Fermi transported [8].

For further reduction of the number of the spin components which are presented explicitly in the MP equations, it is possible to do another step: we can operate only with the three components of the spin tensor  $S^{\lambda\mu}$ which contain only the spatial indices, i.e., $S^{ik}$,  because  the three other components of this tensor by the Mathisson-Pirani condition are expressed as
\begin{equation}
\label{plyatsko:4}
S^{i4}=\frac {u_k}{u_4} S^{ki}.
\end{equation}
The useful form of the MP equation can be obtained if instead of the spin tensor components  $S^{ki}$ the 3-component value  $S_i$ is introduced by definition
\begin{equation}
\label{plyatsko:5}
S_i =\frac{1}{2u_4} \sqrt{-g} \varepsilon_{ikl}S^{kl},
\end{equation}
where  $\varepsilon_{ikl}$ is the spatial Levi-Civita symbol;
the expression of  $S^{ki}$ through $S_i$ is
\begin{equation}
\label{plyatsko:6}
S^{ik}=\frac {u_4}{\sqrt {-g}} \varepsilon^{ikl} S_l.
\end{equation}
Then the transport part of the MP equations  can be  written as [14, 15]
$$
m(\dot u^m + \Gamma^m_{\mu\nu} u^\mu u^\nu)
$$
$$ - 2\frac  {d}{ds}
[|g|^{-1/2} (\dot u^\mu +
\Gamma^\mu_{\pi\rho} u^\pi u^\rho)(u_4 S_{[n} g_{p]\mu}
$$
$$ +
g_{4\mu} S_{[p} u_{n]})] -  |g|^{-1/2}u^\nu(\dot u^\mu
$$
$$
+
\Gamma^\mu_{\pi\rho} u^\pi u^\rho)(u_4\Gamma^m_{i\nu} g_{k\mu} +
2u_i \Gamma^m_{\nu[k} g_{4]\mu})\varepsilon^{ikl}S_l
$$
\begin{equation}
\label{plyatsko:7}
+ \frac12 |g|^{-1/2} u^\pi (u_4 R^m_{\pi ik}
+ 2u_i R^m_{\pi k 4})\varepsilon^{ikl} S_l =0,
\end{equation}
where for the free indices $m$, $n$ and $p$ it is necessary to put the circle
combinations 1, 2, 3; 2, 3, 1; 3, 1, 2, and
the fourth equation of the MP transport part  can be transformed to the
form
$$
m(\dot u^4 + \Gamma^4_{\mu\nu} u^\mu u^\nu)+
\frac {d}{ ds}\left[|g|^{-1/2}  (\dot u^\mu +
\Gamma^\mu_{\pi\rho} u^\pi u^\rho)g_{\mu k}
u_i \varepsilon^{ikl} S_l\right]
$$
$$
-  |g|^{-1/2}u^\nu(\dot u^\mu +
\Gamma^\mu_{\pi\rho} u^\pi u^\rho)(u_4\Gamma^4_{i\nu} g_{k\mu} +
2u_i \Gamma^4_{\nu[k} g_{4]\mu})\varepsilon^{ikl}S_l
$$
\begin{equation}
\label{plyatsko:8}
+ \frac12 |g|^{-1/2} u^\pi (u_4 R^4_{\pi ik}
+ 2u_i R^4_{\pi k 4})\varepsilon^{ikl} S_l =0
\end{equation}
(a dot denotes differentiation with respect to the proper time $s$, and square brackets denote antisymmetrization of  indices).

 The all three independent equations  of the spin part of the MP equations  can be written as [14, 21]
$$
u_{4} \dot {S_i} + 2(\dot u_{[4} u_{i]} -
u^\pi u_\rho \Gamma^\rho_{\pi[4} u_{i]})S_k u^k
$$
\begin{equation}
\label{plyatsko:9}
+ 2S_n \Gamma^n _{\pi [4} u_{i]} u^\pi =0.
\end{equation}

 In practical calculations for the concrete metric, both analytical and numerical, it is often more convenience to use the MP equation representation just through the 3-vector $S_i$. In particular, at a glance, from  (9) we see that the spin part of the MP equations becomes much simpler if the condition $S_i u^i=0$ is satisfied, i.e., when the spin is orthogonal to the particle's trajectory.

According to (1) and (5), there is a simple relationship between $S_i$ and $s_\lambda$:
\begin{equation}
\label{plyatsko:10}
S_i=-s_i+\frac{u_i}{u_4} s_4.
\end{equation}

\subsection{MP equations in comoving tetrad representation}

In many cases, it is important to describe a physical event in different frames of reference. Concerning the motion of a spinning test particle in the gravitational field, it is useful to consider the main properties of this motion from the point of view of an observer moving with the spinning particle, with respect to which the representative point of the particle is at rest.

As usual, the comoving frame of reference are determined by a set of orthogonal tetrads $\lambda^\mu_{(\nu)}$, where  $\lambda^\mu_{(4)}=u^\mu$ and
\begin{equation}
\label{plyatsko:11}
\lambda^\mu_{(\nu)}\lambda^\pi_{(\rho)}g_{\mu\pi}=\eta_{(\nu)(\rho)}, \quad
g_{\mu\nu}=\lambda_\mu^{(\pi)}\lambda_\nu^{(\rho)}\eta_{(\pi)(\rho)}
\end{equation}
(here, and in the following, in contrast to the indices of the global coordinates, the local indices are placed in the parenthesis; $\eta_{(\nu)(\rho)}$ is the Minkowski tensor).

The local components of the spin 4-vector for a comoving observer
satisfy the condition  $s_{(4)} = 0$ [26] . Concerning the spatial components of this vector, without loss in generality, for convince we direct the first vector along the direction of spin, that is we write
\begin{equation}
\label{plyatsko:12}
s_{(1)} \ne 0, \quad s_{(2)} = 0, \quad s_{(3)} = 0.
\quad {s}_{(4)} = 0.
\end{equation}
The MP equations have the constant of motion $S_0^2=s_{(\mu)}s^{(\mu)}$, where $|S_0|$ is the absolute value of the particle spin [26], and according to  (12)   $|s_{(1)}|= |S_0|$.

By relationships  (11) and  (12) for the local components of the spin 3-vector we have
\begin{equation}
\label{plyatsko:13}
S_{(1)}=-s_{(1)} \ne 0, \quad S_{(2)} = 0, \quad S_{(3)} = 0.
\end{equation}
Then according to (10) the global components of this 3-vector are connected with its local components by the relationship
\begin{equation}
\label{plyatsko:14}
\lambda_4^{(4)}\ S_i = \left(\lambda_4^{(4)}\lambda_i^{(1)} -
\lambda_i^{(4)}\lambda_4^{(1)}\right)S_{(1)}.
\end{equation}
It follows from equations (9) that
\begin{equation}
\label{plyatsko:15}
\gamma_{(k)(1)(4)} = 0,
\end{equation}
where $\gamma_{(k)(1)(4)}$ are the Ricci coefficients of rotation.
Because these coefficients are antisymmetric in the first two indices, equation (15) contains the two relationships: $\gamma_{(2)(1)(4)} = 0$ and $\gamma_{(3)(1)(4)} = 0$.
The similar third relationship, $\gamma_{(2)(3)(4)}=0$, can be added if one need exclude the rotation of the second and the third local spatial vectors relative to the direction of the first vector, which in our consideration is connected with the spin orientation. Therefore, we are free to put
\begin{equation}
\label{plyatsko:16}
\gamma_{(i)(k)(4)} = 0.
\end{equation}
Equation  (16) is the known condition of the Fermi transport for the local orthonormal vectors.

By direct calculations, which are quite simple but rather lengthy, it follows from equations (7)  that
\begin{equation}
\label{plyatsko:17}
m\gamma_{(1)(4)(4)} = S_{(1)} R_{(1)(4)(2)(3)} ,
\end{equation}
$$
m\gamma_{(2)(4)(4)} = S_{(1)}(R_{(2)(4)(2)(3)}
$$
\begin{equation}
\label{plyatsko:18}
- \dot \gamma_{(3)(4)(4)}
- \gamma_{(2)(4)(4)} \gamma_{(2)(3)(4)}),
\end{equation}
$$
m\gamma_{(3)(4)(4)} =S_{(1)}(R_{(3)(4)(2)(3)}
$$
\begin{equation}
\label{plyatsko:19}
+ \dot \gamma_{(2)(4)(4)}
- \gamma_{(3)(4)(4)} \gamma_{(2)(3)(4)}).
\end{equation}
Here, as for the spin part of the MP equations, the first local spatial vector is determined by the spin orientation.

It is known that the value $\gamma^{(i)}_{~(4)(4)}$ is the dynamical characteristic of the reference frame, namely, its acceleration.

In the linear spin approximation  instead of equations (17)--(19) we have the more simple equations
\begin{equation}
\label{plyatsko:20}
m\gamma_{(i)(4)(4)} = -S_{(1)} R_{(1)(4)(2)(3)}.
\end{equation}
That is, according to (20) the force of the spin-gravity interaction, as estimated by the comoving observer in the linear spin approximation, is fully
determined by the local components of the Riemann tensor. We shell analyze expression (20) in some concrete cases of  spinning particle motions in the gravitational field.

 \section{Highly relativistic solutions \\ of MP equations \\ in Schwarzschild's background}

We begin this section from consideration of a physical situation which is not connected directly with the MP equations and, in a certain sense, is similar to the known one in electrodynamics, when the electromagnetic field of a moving electric charge is under consideration.
Now we are interested in the gravitational field of a moving mass. More exactly, let us analyze the expressions for Riemann's tensor components as evaluated by an observer which is moving with any
velocity relative to Schwarzschild's mass.

\subsection{Gravitational field of a moving Schwarzschild's mass}

 As usual, to describe a moving observer we can use the corresponding set of orthogonal tetrads $\lambda^\mu_{~(\nu)}$. Following many papers where the deeper analogies between gravitation and electromagnetism are under investigations we consider the gravitoelectric
 $E_{(k)}^{(i)}$ and gravitomagnetic   $B_{(k)}^{(i)}$ components of the gravitational field. For example, in [39] these components are determined as
\begin{equation}\label{plyatsko:21}
E_{(k)}^{(i)}=R^{(i)(4)}_{}{}{}{}{}{}_{(k)(4)},
\end{equation}
\begin{equation}\label{plyatsko:22}
B_{(k)}^{(i)}=-\frac12 R^{(i)(4)}_{}{}{}{}{}{}_{(m)(n)}
\varepsilon^{(m)(n)}_{}{}{}{}{}{}_{(k)}.
\end{equation}
Here our consideration is restricted  by the gravitomagnetic components only for the observer which is moving  relative to  Schwarzschild's mass.
 The space local vectors can be oriented as:
(1) The first vector is orthogonal to the plane that is determined by the instantaneous direction of the observer motion relative to the Schwarzschild mass and the radial direction; (2) The second vector is directed along the direction of the motion of the observer.
Then using the standard Schwarzschild coordinates we obtain the nonzero components of the gravitomagnetric field
\begin{equation}\label{plyatsko:23}
B^{(1)}_{(2)}=B^{(2)}_{(1)}=
\frac{3Mu_\parallel u_\perp}
{r^3\sqrt{u_4u^4-1}}\left(1-\frac{2M}{r}\right)^{-1/2},
\end{equation}
\begin{equation}\label{plyatsko:24}
B^{(1)}_{(3)}=B^{(3)}_{(1)}=
\frac{3M u_\perp^2 u^4}
{r^3\sqrt{u_4u^4-1}}\left(1-\frac{2M}{r}\right)^{1/2},
\end{equation}
where  $u_\parallel\equiv dr/ds$ and  $u_\perp\equiv rd\varphi/ds $ are the radial and tangential components of the observer's 4-velocity, and  $M$ is Schwarzschild's mass.

Let us analyze expressions (20) at different velocities of the observer relative to Schwarzschild's mass. Note that the gravitomagnetic components  are nonzero only at $u_\perp\ne 0$. That is, here the situation is similar to the known one in the electrodynamic for the components of the magnetic field of a moving electric charge. Then it is easy to see that the components (20) significantly depend on the velocity of an observer relative to the Schwarzschild mass. Indeed, at
$ |u_\perp|\ll 1 $, $ |u_\parallel|\ll 1 $,
 the common term $3M/r^3$ in expressions (20) is multiplied on the corresponding small values. Whereas in the highly relativistic case, for $ |u_\perp|\gg 1 $, this term is multiplied by the large (as compare to 1) values and then we have
\begin{equation}\label{plyatsko:25}
B^{(1)}_{(2)}=B^{(2)}_{(1)} \sim \frac{3M}{r^3}\gamma,\quad
B^{(1)}_{(3)}=B^{(3)}_{(1)} \sim \frac{3M}{r^3}\gamma^2,
\end{equation}
where $ \gamma $ is the relativistic Lorentz factor as calculated by the particle velocity relative to the Schwarzschild mass.

\subsection{Highly relativistic spin-orbit acceleration \\ as measured by a comoving observer}

 What physical effects are caused by the gravitomagnetic components (23)--(25)?
To answer this question, let us consider equation (20) which can be written as
\begin{equation}\label{plyatsko:26}
a_{(i)} = -\frac {S_{(1)}}{m} R_{(i)(4)(2)(3)},
\end{equation}
where  $a_{(i)}$ are the local components of the particle 3-acceleration relative to geodesic free fall as measured by the comoving observer;  $S_{(1)}$ is the single nonzero component of the particle spin, i.e., here the first local space vector is oriented along the spin.

In the concrete case of particle motion in Schwarzschild's background, when the particle's spin is orthogonal to the plane determined by the direction of particle motion and the radial direction, by (22) and (26) we have
\begin{equation}\label{plyatsko:27}
a_{(i)}=-\frac{S_{(1)}}{m}B^{(1)}_{(i)},
\end{equation}
where the nonzero values of $ B^{(1)}_{(i)} $ come from (23) and (24) . According to
(25), the acceleration components (27) depend, in the case of highly relativistic nonradial motions, on $\gamma $ such that  $ a_{(2)} \sim \gamma $, $ a_{(3)} \sim \gamma^2$. The component $ a_{(1)}$ remain equal to zero at any velocity. The absolute value of the 3-acceleration is proportional to $ \gamma^2 $.

So, according to the MP equations, from the point of view of the observer comoving with a spinning particle in  Schwarzschild's background, the spin-gravity interaction is much greater at the highly relativistic particle's velocity than at the low velocity. This interaction has the clear feature of the spin-orbit interaction. However, it is interesting to estimate the effects of this  interaction for another observer, e.g., which is at rest relative to the Schwarzschild mass.

\subsection{Exact MP equations in Schwarzschild's background \\ after using constants of motion}

Due to the symmetry of Schwarzschild's and  Kerr's metric here the MP equations have the constants of motion: the particle's energy $E$ and the projection of its angular momentum $J_z$ [5, 40--42]. It is known that in the case of the geodesic equations the analogous constants of motion were effectively used for analyzing possible orbits of a spinless particle in these metrics [43]. Namely, by the constants of energy and angular
momentum the standard form of the geodesic equations, which are
the differential equations of the second order by the coordinates,
can be reduced to the differential equations of the first order.
Naturally, it is interesting to apply the similar procedure to the
exact MP equations. However, in contrast to
the geodesic equations, the exact MP equations at the Mathisson-Pirani condition contain the third derivatives of the coordinates.

The corresponding explicit form of the MP equations in Kerr's background is presented in [21]. Here we write these equations for Schwarzschild's background. As in [21], we  use  the set of the 11 dimensionless quantities $y_i$ where by definition
$$
\quad y_1=\frac{r}{M},\quad y_2=\theta,\quad y_3=\varphi, \quad
y_4=\frac{t}{M},
$$
$$
y_5=u^1,\quad y_6=Mu^2,\quad y_7=Mu^3,\quad y_8=u^4,
$$
\begin{equation}\label{plyatsko:28}
    y_9=\frac{S_1}{mM},\quad y_{10}=\frac{S_2}{mM^2},\quad
    y_{11}=\frac{S_3}{mM^2}.
\end{equation}
 In addition, we introduce the dimensionless quantities
connected with the particle's proper time $s$ and the constants of
motion $E$ , $J_z$:
\begin{equation}\label{plyatsko:29}
    x=\frac{s}{M}, \quad \hat E=\frac{E}{m},\quad
    \hat J=\frac{J_z}{mM}.
\end{equation}
Some quantities from (28) satisfy the four simple equations
\begin{equation}\label{plyatsko:30}
\dot y_1=y_5, \quad \dot y_2=y_6, \quad \dot y_3=y_7, \quad \dot
y_4=y_8,
\end{equation}
here and in the following a dot denotes the usual derivative with
respect to $x$.  Other seven equations for the 11  functions $y_i$ can be written as
\begin{equation}\label{plyatsko:31}
   \dot y_5 = A_1,\quad \dot y_6 = A_2,\quad
\dot y_7 = A_3,\quad \dot y_8 = A_4,
\end{equation}
\begin{equation}\label{plyatsko:32}
   \dot y_9 = A_5,\quad \dot y_{10} = A_6,\quad
\dot y_{11} = A_7,
\end{equation}
where
\[
 A_1=A+
\left(y_1y_6^2+y_1y_7^2\sin^2{y_2}-\frac{y_8^2}{y_1^2}\right)q
+\frac{y_5^2}{y_1^2q},
\]
\[
A_2=B-\frac{2}{y_1}y_5y_6+y_7^2\sin{y_2}\cos{y_2},
\]
\[
 A_3=\left(-q^{-1}y_5y_{10}+y_1^2y_6y_9\right)^{-1}\times
[q^{-1}(-y_7y_{10}\sin^2{y_2}\]
\[+y_{6}y_{11})A+
\left(-q^{-1}y_5y_{11}
+y_1^2y_7y_9\sin^2{y_2}\right)B\]
\[-y_8\sin{y_2}+\frac{\hat E}{\varepsilon_0 q}\sin{y_2}+\frac{1}{y_1^2 q}
(y_7y_{10}\sin^2{y_2}\]
\[y_6y_{11})]\sin^{-2}{y_2}-\frac{2}{y_1} y_5y_7-2y_6y_7\cot{y_2},
\]
\[A_4=y_8\left(-q^{-1}y_5y_{10}+y_1^2y_6y_9\right)^{-1}\]
\[\times
[-q^{-1}y_{10}A+y_1^2y_9B- y_1^2\frac{y_7}{y_8q}\sin{y_2}\]
\[+\frac{\hat J_z}{\varepsilon_0\sin{y_2}}+
\frac{y_{10}}{y_1}-y_9\cot{y_2}]-\frac{2}{y_1^2}y_5y_8q^{-1},\]

\[A=\varepsilon_0^{-2}q^{-1}\left(q^{-1}
y_5y_{10}-y_1^2y_6y_9\right)^{-1}\]
\[\times\Big[[y_7y_8q(-y_5y_{10}y_1^{-2}+q
y_9(y_6y_{10}+y_7y_{11}))\]
\[+\frac{y_8 y_{11} q}{y_1^2\sin^2{y_2}}
(qy_9+y_5(y_5y_9+y_6y_{10}))]\]
\[
\times [y_1^2\frac{\hat E}{q\varepsilon_0y_8}\sin{y_2}+
\frac{1}{qy_8^2}(y_7y_{10}\sin^2{y_2}-y_6y_{11})]\]
\[+\frac{q y_8}{y_1^2}[
y_5y_{10}^2+\frac{y_5y_{11}^2}{\sin^2{y_2}}-qy_1^2y_9(y_6y_{10}
+y_{7}y_{11})]\]
\[\times\left(\frac{\hat J_z}{\varepsilon_0\sin{y_2}}+\frac{q y_8 y_{10}}{y_1}-y_9\cot{y_2}\right)\Big]\]
\[+ \frac{qy_8}{\varepsilon_0^2\sin{y_2}}[y_9+q^{-1}y_5(y_5y_9+
y_6y_{10}+y_7y_{11})]\]
\[\times\Big[\frac{6y_9}{y_1^3}q(
-y_7y_{10}\sin^2{y_2}+y_6y_{11})\]
\[ -y_{11}(q^{-1}
y_5y_{10}-y_1^2y_6y_9)^{-1}\Big],
\]

\[B=y_1^{-2}\left[y_9+q^{-1}y_5(y_5y_9+y_6y_{10}+y_7y_{11})\right]
^{-1}\]
\[\times\Big[ q^{-1}
(y_{10}+y_1^2y_6(y_5y_9+y_6y_{10}+y_7y_{11}))A \]
\[+\frac{\hat E}{q\varepsilon_0}
y_1^2y_7\sin{y_2}+q^{-1}y_7 (y_7y_{10}\sin^2{y_2}-y_6y_{11})\]
\[
-y_8 \left(\frac{\hat J_z}{\varepsilon_0\sin{y_2}}+q y_8y_{10}y^{-1}_1-qy_8y_9\cot{y_2}\right)\Big].
\]

\[A_5=qy_8\frac{y_6y_{10}+y_7y_{11}}{y_1} -qy_8(y_5y_9+y_6y_{10}
+y_7y_{11})
\]
\[\times\left[\left(A_1-\frac{y_5A_4}{y_8}-\frac{y_5^2}{qy_1^2}\right)q^{-1}
-y_1y_6^2-y_1y_7^2\sin^2{y_2}+\frac{y_8^2}{y_1^2}\right],\]

\[A_6=-qy_6y_8y_9(y_1-3)+\frac{qy_5y_8 y_{10}}{y_1}
\]\[+qy_7y_8y_{11}\cot{y_2}-qy_8(y_5y_9+y_6y_{10}+y_7y_{11})\]
\[\times\Big[y_1^2A_2-\frac{y_1^2y_6}{y_8}A_4+2y_5y_6
\left(y_1-q^{-1}\right)-y_1^2y_7^2\cos{y_2}\sin{y_2}\Big],\]

\[A_7=\frac{qy_5y_8y_{11}}{y_1}-qy_7y_8y_9(y_1-3)
\sin^2{y_2}
\]
\[+qy_6y_8y_{11}\cot{y_2}-qy_7y_8y_{10}\cos{y_2}\sin{y_2}
\]
\[- qy_8(y_5y_9+y_6y_{10}+y_7y_{11})\times
\Big[y_1^2A_3\sin^2{y_2}-\frac{y_1^2y_7}{y_8}A_4\sin^2{y_2}\]
\begin{equation}\label{plyatsko:33}
+2y_5y_7(y_1-q^{-1})\sin^2{y_2}+2y_1^2y_6y_7\cos{y_2}\sin{y_2}\Big].
\end{equation}
In all expressions  (33) we note
\[
q=1-\frac{2}{y_1}
\]
and
\begin{equation}\label{plyatsko:34}
\varepsilon_0\equiv\frac{|S_0|}{mM}.
\end{equation}
We stress that while dealing with the MP equations in Schwarzschild's or Kerr's background the condition for a spinning test particle
\begin{equation}\label{plyatsko:35}
 \frac{|S_0|}{mr}\equiv\varepsilon\ll 1
\end{equation}
must be taken into account [10].

\subsection{Highly relativistic circular equatorial orbits \\ of a spinning particle}

Equations (31) and (32) can be used for  description of all possible types of the spinning particle motions in Schwarzschild's background according to the MP equations under the Mathisson-Pirani condition. We begin from considering the important partial solutions which describe the circular orbits of a spinning particle in the equatorial plane $\theta=\pi/2$ of Schwarzschild's metric when the spin is orthogonal to this plane. Then all the MP equations are reduced to the single algebraic equation, where among the values $y_i$ only  $y_1$,  $y_7$, and  $y_8$ are
present [22]:
$$
y_7^3(y_1-3)^2y_8y_1^2\varepsilon_0-y_7^2(y_1-3)y_1^3
$$
\begin{equation}\label{plyatsko:36}
+y_7(2y_1-3)\varepsilon_0 y_8+y_1=0
\end{equation}
(without any loss in generality, this equation is written for the
orientation of the particle's spin when $S_2\equiv S_{\theta} >0$).
Per relationship $u_\mu u^\mu =1$ the value $y_8$ can be expressed through $y_7$:
\begin{equation}\label{plyatsko:37}
y_8=\left(1-\frac{2}{y_1}\right)^{-1/2}\sqrt{1+y_1^2 y_7^2}.
\end{equation}
In the limiting transition to the spinless particle ($\varepsilon_0=0$) we have from (36) the result known from the geodesic equations.

Equation (36) with (37) determine the region of existence of the circular orbits and the dependence of the particle's angular velocity, which in notation (28) correspond to $y_7$, on the radial coordinate. Naturally, it is interesting to compare the situations with the spinless and spinning particle. The
known result is that by the geodesic equations these orbits are
allowed only for $r>3M$  and the highly
relativistic circular orbits exist only for $r=3M(1+\delta)$,
where $0<\delta \ll 1$. In this context we point out that equation (36)
admits the highly relativistic circular orbit of a spinning particle with $r=3M$, i.e., $y_1=3$. Indeed, it is easy to calculate the single real root
$y_7$ of equation (36) in the simple analytical form
\begin{equation}\label{plyatsko:38}
y_7=-\frac{3^{-3/4}}{\sqrt{\varepsilon_0}}(1+O(\varepsilon_0))
\end{equation}
(here we take into account that by (35) it is necessary $\varepsilon_0\ll 1$). Then the absolute value of the particle orbital 4-velocity, which corresponds to (38), in the main spin approximation is equal to
$3^{1/4}/\sqrt{\varepsilon_0}$. The same value has the relativistic Lorentz $\gamma$-factor of the spinning particle as calculated by an observer which is at rest relative to Schwarzschild's mass. Hence $\gamma^2\gg 1,$
i.e., for the motion on the circular orbit with  $r=3M$ the velocity of a spinning particle must be ultrarelativistic, the more ultrarelativistic the smaller is the ratio of spin to mass of the particle.

It is known that by the geodesic equations a spinless particle with nonzero mass of any velocity close to the velocity of light, starting in the tangential direction from the position $r=3M$, will fall on the horizon surface within a finite proper time. On the other hand, a spinning particle can remain indefinitely on the circular orbit with $r=3M$ due to the highly relativistic spin-gravity coupling if condition (38) is satisfied. It means that in this concrete situation the spin-gravity coupling acts as a strong repulsive force which compensates the usual (geodesic) attraction. By the way, from the point of view of an observer comoving with the spinning particle the value of this force is the same order as the formally calculated value of the Newtonian gravitational force for the masses $M$ and $m$ at distance $r=3M$ [16].

Certainly, the picture described for the spinning particle on the circular orbit with $r=3M$ holds in the ideal case only, when perturbations are neglected.
Some cases with perturbations (noncircular motions) we shall consider below.

We stress the important feature of the partial circular solution of the MP equations with (38): this solution is common for the exact MP equations and their linear spin approximation and, as a result, both for the Mathisson-Pirani and  Tulczyjew-Dixon condition [14, 17].

Let us consider other circular orbits of a spinning particle in Schwarz\-schild's background which are described by equation  (38). The detailed analysis is presented in [22] and here we show some graphs which illustrate the domain of existence of the highly relativistic circular orbits and the dependence of the $\gamma$-factor, which is necessary for realizing these orbits, on the radial coordinate. Figure 1 describes the orbits with the negative values of the particle orbital velocity ($y_7<0$) at $S_\theta>0$, i.e., the signs of these values are the same as in the case of the highly relativistic circular orbit with  $y_1=3$ and $y_7$ from (38).
(In these figures below we put $\varepsilon_0=10^{-2}$).
The orbits corresponding to Fig. 1 exist in the space region $2M<r<3M$ and, similarly to  the circular orbit with  $y_1=3$, these orbits are caused by the significant repulsive action of the highly relativistic spin-gravity coupling. However, in contrast to the orbit  with  $y_1=3$, which is fully determined by the linear spin terms, for the orbits in Fig. 1  the nonlinear terms are important. In this context Fig. 2 shows the difference in the description of the highly relativistic circular orbits of a spinning particle in the small neighborhood of $y_1=3$, that is when $y_1=3+\delta$, $\delta<0$, $|\delta|\ll 1$, by the exact MP equations and their linear spin approximation.

According to the solutions of equation  (36) for different $y_1$, the highly relativistic orbits of a spinning particle in Schwarzschild's background exist for $r>3M$ as well. However, in contrast to the orbits from $2M<r<3M$, they are caused by the significant attractive action of the spin-gravity coupling, and for realizing these orbits the sign of the particle orbital velocity must be positive ($d\varphi/ds>0$) for $S_\theta>0$. Figure 3 describes the small space domain, where  $y_1=3+\delta$, $0<\delta\ll 1$,  and Fig. 4 corresponds to the region for the large values of $y_1$.
There is an essential difference between the upper and lower part of
the solid curve in Fig. 3. Namely, the last  curve is close to the
dashed line for $\delta>0.00631$ and both these curves tend to the
geodesic line if $\delta$ is growing, whereas the upper part of the
solid curve in Fig. 3  significantly differs from the geodesic line.
The dependence of the $\gamma$ on $r/M$ for this case on the
interval from 3.02  is presented in Fig. 4, and  for $r\gg 3M$ the value $\gamma$ is proportional
to $\sqrt{r}$. It means that for the motion on a circular orbit with
$r\gg 3M$ the particle must posses much higher orbital velocity than
in the case of the motion on a circular orbit near $r=3M$.
\begin{figure}
\centering
\includegraphics[width=5cm]{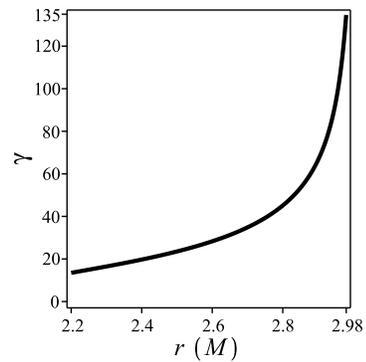}
\caption{\label{plyatsko_fig:1} Lorentz factor vs $r$ for the highly
relativistic circular orbits with $d\varphi/ds<0$ of the spinning
particle beyond the small neighborhood of $r=3M$ according to the
exact MP equations.}
\end{figure}

\begin{figure}
\centering
\includegraphics[width=5cm]{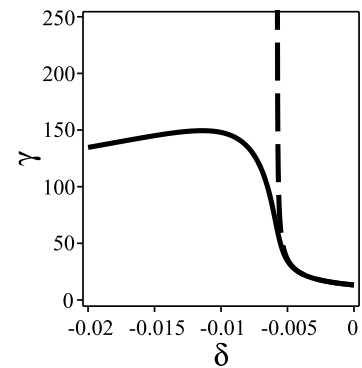}
\caption{\label{plyatsko_fig:2} Lorentz factor vs $\delta$ for the highly
relativistic circular orbits with $d\varphi/ds<0$ of the spinning
particle in the small neighborhood of $r=3M$ according to the
exact MP equations  (solid
line) and their linear spin approximation (dashed line).}
\end{figure}

\begin{figure}
\centering
\includegraphics[width=5cm]{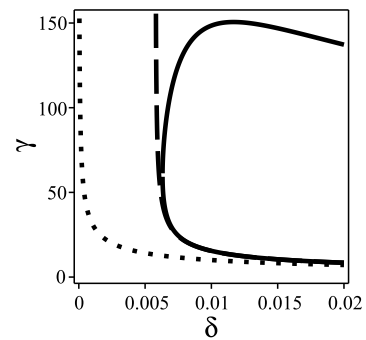}
\caption{\label{plyatsko_fig:3} Dependence of the Lorentz factor on $\delta>0$
for  the  highly relativistic circular orbits with $d\varphi/ds>0$
of the spinning particle in the small neighborhood of $r=3M$
according to the exact MP equations and their linear spin approximation (solid line) and their linear spin approximation (dashed
line). The dotted line corresponds to the geodesic circular
orbits.}
\end{figure}

\begin{figure}
\centering
\includegraphics[width=5cm]{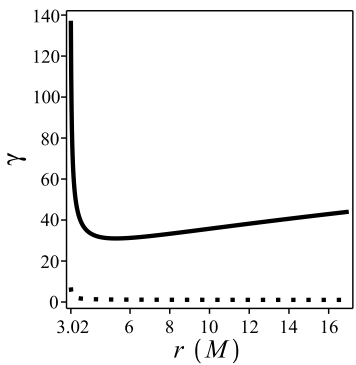}
\caption{\label{plyatsko_fig:4} Dependence of the Lorentz factor on $r$ for the
highly relativistic circular orbits with $d\varphi/ds>0$ of the
spinning particle beyond the small neighborhood of $r=3M$ by the
exact MP equations  (solid
line). The dotted line corresponds to the geodesic circular
orbits.}
\end{figure}

For deeper understanding of  physics of the highly relativistic
circular orbits of a spinning particle in Schwarzschild's
background, let us estimate the values of the particle's energy $E$
on these orbits. It is not difficult to calculate  the values for $E$  in the case of
the circular orbits in the Schwarzschild background as [22]
\begin{equation}\label{plyatsko:39}
E=m\left[\left(1-\frac{2}{y_1}\right) y_8 - \varepsilon_0 y_1(y_1-3)y_7^3\right].
\end{equation}
It is easy to check that by  (39)
the energy of the spinning particle on the above considered
highly relativistic circular orbits is positive and much less than
the energy of the spinless particle on the corresponding geodesic
circular orbits. For example, the energy of the spinless particle
on the circular orbits with $r>3M$ tends to $\infty$ if $r\to 3M$,
whereas according to  (36)  the energy of the spinning particle is
finite for its circular orbits with any $r$, including $r=3M$. In
the case of the highly relativistic circular orbits of the
spinning particle beyond the small neighborhood of $3M$ it follows from  (39) that
\begin{equation}\label{plyatsko:40}
E=m\frac{\sqrt{\varepsilon_0}}{\sqrt{y_1}}
\left(1-\frac{2}{y_1}\right)^{1/4}\left|1-\frac{3}{y_1}\right|^{-3/2}
\left(1-\frac{3}{y_1}+\frac{3}{y_1^2}\right).
\end{equation}
Hence, by (40)  we have $E^2\ll m^2$ (note that the
right-hand side of equation  (40) is positive for all values $y_1$
beyond the horizon surface). That is, in this sense one can draw a
conclusion concerning the strong binding energy for those orbits
which is caused by the interaction of  spin with the
gravitational field.

\subsection{Circular nonequatorial orbits}

It is not difficult to verify that the MP equations in the standard coordinates of the Schwarzschild metric have the partial solution with [14]
\begin{equation}
\label{plyatsko:41}
r=const\ne 0, \quad \theta=const\ne 0, \pi/2, \pi,
\end{equation}
\begin{equation}
\label{plyatsko:42}
u^3=\frac{d\varphi}{ds}=const\ne 0, \quad u^4=\frac{dt}{ds}=const\ne 0,
\end{equation}
\begin{equation}
\label{plyatsko:43}
S_r=const\ne 0, quad S_\theta=const\ne 0, \quad S_\varphi=0,
\end{equation}
\begin{equation}
\label{plyatsko:44}
S_r\left(1-\frac{3M}{r}\right)+S_{\theta}\frac{\cos\theta}{r\sin\theta}=0.
\end{equation}
Expressions (41)--(44) describe the nonequatorial circular orbits of a spinning particle with the constant orbital velocity when the spin lays in the meridional plane when the particle orbits the Schwarzschild mass. The necessary particle's velocity for motions on these orbits is determined by
\begin{equation}
\label{plyatsko:45}
u^3=-\frac{K^{1/4}\left(1-\frac{2M}{r}\right)^{-1/4}}{r\sqrt{6\varepsilon |\sin\theta}|sign(S_\theta \sin\theta)}\left[1+O(\varepsilon)\right],
\end{equation}
where we note
\begin{equation}
\label{plyatsko:46}
K=1+\left(1-\frac{2M}{r}\right)\left(1-\frac{3M}{r}\right)^{-2}\frac{\cos^2\theta}{\sin^2\theta}>0,
\end{equation}
and $\varepsilon$ is  the small value as determined by  (35).
The dependence  of the angle $\theta$ on $r$ is
$$
\sin^2\theta=\left(1-\frac{2M}{r}\right)[\frac Mr \left(4-\frac{9M}{r}\right)
$$
\begin{equation}
\label{plyatsko:47}
-6\left(1-\frac{2M}{r}\right)
\left(1-\frac{3M}{r}\right)]^{-1}(1+O(\varepsilon)).
\end{equation}
For the nonequatorial circular orbits it is necessary $0<\sin^2\theta<1$. It follows from  (47) that this condition is satisfied if and only if
\begin{equation}
\label{plyatsko:48}
\frac{15}{7} M\left(1+O(\varepsilon)\right)<r<3M\left(1+O(\varepsilon)\right).
\end{equation}

Therefore, relationships (45)--(48) give the necessary and sufficient conditions for the motion of a spinning test particle on the nonequatorial circular orbits in the Schwarzschild background.  According to (48), the space region of existence of these orbits by the radial coordinate is determined (without small correction of the order $\varepsilon$) as
\begin{equation}
\label{plyatsko:49}
\frac{15}{7} M<r<3M.
\end{equation}
 It follows from (47) that the minimum value of
$\sin^2\theta$ is achieved at
$$
r=\frac{5\sqrt 5}{3\sqrt 5 -1}\approx 2.35M,
$$
and
$$
\sin^2\theta|_{min}\approx 0.465, \quad \theta_{min}\approx43^{\circ}.
$$
Whereas at $r=2.35M$ and $r=2.5M$ we have from (47) the common values $\sin^2\theta=0.5$ and $\theta=45^{\circ}$.

We stress that for the realization of the considered nonequatorial circular orbits a spinning particle  must posses the highly relativistic velocity, in the sense that
 $\gamma^2\approx 1/\varepsilon\gg 1$.

Concerning the correlation between the direction of the particle orbital motion and its spin orientation we conclude from (45) that, for example, when $\sin\theta>0$ ($0<\theta<\pi/2$) the signs of $S_\theta$ and $u^3=d\varphi/ds$ are opposite.

The considered circular orbits one can called  {\it hovering}, because a spinning particle hovers above the Schwarzschild mass under the angle
$\alpha_{hover}=\pi/2-\theta$. According to (44) if $r$ is in the small neighborhood of the value $3M$, the relationship $|S_r|\ll |S_\theta|$ takes place, i.e., in this case the particle's spin practically lays in the equatorial plane and by (47) $\alpha_{hover}$ is close to
$0^{\circ}$. Whereas if $r$ is in the small neighborhood of the value $15M/7$, the relationship $|S_\theta|\ll |S_r|$ is valid, i.e., here the spin is practically orthogonal to the equatorial plane and $\alpha_{hover}$ is close to $0^{\circ}$. Then it follows from (45) that if $r\to 15M/7$ the expression for the particle's orbital velocity pass the corresponding expression for the highly relativistic equatorial circular orbit with $r=15M/7$.

We stress that equations (44)--(47) were obtained from the exact MP equations under the Mathisson-Pirani supplementary condition. The MP equations in the linear spin approximation have the solutions which describe the nonequatorial highly relativistic circular orbits of a spinning particle in Schwarzschild's background as well. However, these solutions exist only in the small neighborhood of the radial coordinate $r=3M$ under the small hovering angle.

\subsection{Highly relativistic noncircular orbits}

While investigating the highly relativistic noncircular orbits of a spinning particle in the Schwarzschild background it is necessary to integrate the set
of the differential equations (31)--(33). Note that this system  contains the two parameters $\hat E$ and  $\hat J$  proportional to the constants of the particle's
motion: the energy and angular momentum. By choosing different
values of these parameters for the fixed initial values of $y_i$ one
can describe the motions of different centers of mass. To describe
the proper center of mass of a spinning particle in the
Schwarzschild background, the method of separation of the
corresponding solutions of the exact MP equations was proposed and realized in explicit relationships for the cases when the initial radial component of the particle velocity is much less than the tangential one [18]. The corresponding motions are considered and discussed in detailed in [18, 22]. Here we present
three typical figures which illustrate some important features of the highly relativistic noncircular nonequatorial motions of a spinning particle in Schwarzschild's background (more illustrations are in [20]). Figures 5 and 6 show the situations when the highly relativistic spin-gravity coupling causes the significant attractive and repulsive action correspondingly (the circles with the radius 2 in these figures correspond to the horizon line).

Figure 7 illustrates the cases of the nonequatorial motions when the particle's spin is not orthogonal to the equatorial plane $\theta=\pi/2$. More exactly, this figure describes $\theta$-oscillations for different values of the inclination angle of the spin to the equatorial plane due to $S_r\ne 0$ with $S_\varphi=0$. All graphs in Fig. 7 are interrupted beyond the domain of validity of the linear spin approximation. Within the time of this approximation validity, the period of the $\theta$-oscillations coincides with the period of the
particle's revolution by $\varphi$. Whereas on this interval the
value of the spin component $S_1$ is $const$, just as the
components $S_2$, $S_3$, and the coordinate $r$ . This situation differs from the
corresponding case of the circular motions of a spinning particle
that are not highly relativistic. Indeed, then the nonzero radial
spin component is not $const$ but oscillates with the period of
the particle's revolution by $\varphi$. Besides, in the last case
the mean level of $\theta$ coincides with $90^{\circ}$, whereas in
Fig. 7 the mean values of $\theta$ are above $90^{\circ}$.
According to this figure the amplitude of the $\theta$-oscillation
increases with the inclination angle. However, $\theta -
90^{\circ}$ is small even for the inclination angle that is equal
to $90^{\circ}$.
\begin{figure}
\centering
\includegraphics[width=5cm]{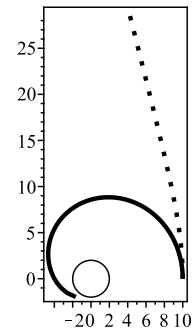}
\caption{\label{plyatsko_fig:5} Trajectories in the polar coordinates of the spinning (solid line) and spinless particle (dotted line) with the same initial values of their 4-velocity components: the tangential and radial velocities are equal to $\approx 35$ and $10^{-2}$ correspondingly. The initial value of the radial coordinate is equal to $10M$, $\varepsilon_0=10^{-2}$.}
\end{figure}

\begin{figure}
\centering
\includegraphics[width=5cm]{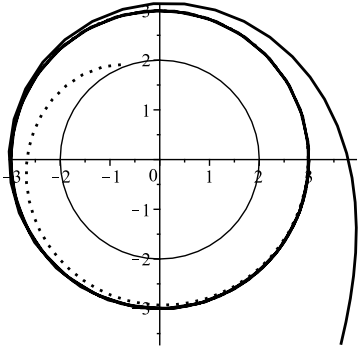}
\caption{\label{plyatsko_fig:6} Trajectories  of the spinning (solid line) and spinless particle (dotted line) with the same initial value of  the tangential velocity,  which is equal to the spinning particle velocity on the circular orbit with $r=3M$, and with much less radial velocity of order  $10^{-7}$. The initial value of the radial coordinate is equal to $3M$ and $\varepsilon_0=10^{-4}$.}
\end{figure}

\begin{figure}
\centering
\includegraphics[width=5cm]{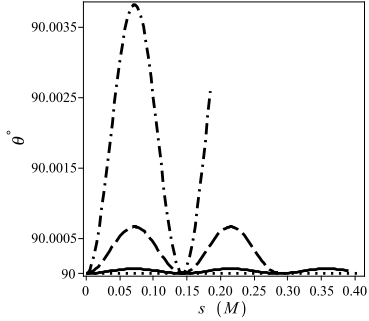}
\caption{\label{plyatsko_fig:7} Graphs of the angle $\theta$ vs proper time
for the inclination angle $0^{\circ}$ (horizontal line
$\theta=90^{\circ})$, $1^{\circ}$ (solid line), $10^{\circ}$ (dash
line), and $90^{\circ}$ (dash and dot line) at
$\varepsilon_0=10^{-4}$.}
\end{figure}

\subsection{A restriction on using the Tulczyjew-Dixon condition}

Now we consider some interesting situation which arise when the highly relativistic solutions of the MP equations under the Tulczyjew-Dixon condition are studied [21].
Let us write the main relationships following from the MP
equations at Tulczyjew-Dixon condition. The mass of a spinning
particle $\mu$ is defined as
\begin{equation}\label{plyatsko:50}
\mu=\sqrt{P_\lambda P^\lambda}
\end{equation}
and $\mu$ is the integral of motion. The quantity
$V^\lambda$ is the normalized momentum, where by definition
\begin{equation}\label{plyatsko:51}
V^\lambda=\frac{P^\lambda}{\mu}.
\end{equation}
 As the normalized quantities $u^\lambda$ and
$V^\lambda$ satisfy the relationships
\begin{equation}\label{plyatsko:52}
u_\lambda u^\lambda=1,\quad V_\lambda V^\lambda=1.
\end{equation}
There is the important relationship between $u^\lambda$ and
$V^\lambda$ [40--42]:
\begin{equation}\label{plyatsko:53}
    u^{\lambda}=N\left[V^\lambda+\frac{1}{2\mu^2\Delta}
    S^{\lambda\nu}V^{\pi}R_{\nu\pi\rho\sigma}S^{\rho\sigma}\right],
\end{equation}
where
\begin{equation}\label{plyatsko:54}
\Delta=1+\frac{1}{4\mu^2}R_{\lambda\pi\rho\sigma}S^{\lambda\pi}S^{\rho\sigma}.
\end{equation}

Now our aim is to consider the explicit form of expression (53)
for the concrete case of the Schwarzschild metric, for the
particle motion in the plane $\theta=\pi/2$ when spin is
orthogonal to this plane (we use the standard Schwarzschild
coordinates). Then we have
\begin{equation}\label{plyatsko:55}
u^2=0,\quad u^1\ne 0, \quad u^3\ne 0, \quad u^4\ne 0,
\end{equation}
\begin{equation}\label{plyatsko:56}
S^{12}=0,\quad S^{23}=0,\quad S^{13}\ne 0.
\end{equation}
In addition to (56) by Tulczyjew-Dixon condition  we write
\begin{equation}\label{plyatsko:57}
 S^{14}=-\frac{P_3}{P_4}S^{13},\quad S^{24}=0, \quad S^{34}=\frac{P_1}{P_4}S^{13}.
\end{equation}
Using  corresponding expressions for the
Riemann tensor in the Schwarzschild metric, from (53) we obtain
\[
u^1=NV^1\left(1+\frac{3M}{r^3}V_3
V^3\frac{S_0^2}{\mu^2\Delta}\right), \quad u^2=V^2=0,
\]
\[
u^3=NV^3\left[1+\frac{3M}{r^3}(V_3
V^3-1)\frac{S_0^2}{\mu^2\Delta}\right],
\]
\begin{equation}\label{plyatsko:58}
u^4=NV^4\left(1+\frac{3M}{r^3}V_3
V^3\frac{S_0^2}{\mu^2\Delta}\right),
\end{equation}
 According to (54)
we write the expression for $\Delta$ as
\begin{equation}\label{plyatsko:59}
\Delta=1+\frac{S_0^2 M}{\mu^2 r^3}(1-3V_3 V^3)
\end{equation}
(the quantity $M$  is the mass of a Schwarzschild
source). Inserting (59) into (58) we get
\[
u^1=\frac{NV^1}{\Delta}\left(1+\frac{S_0^2 M}{m^2 r^3}\right),
\]
\[u^3=\frac{NV^3}{\Delta}\left(1-2\frac{S_0^2 M}{m^2 r^3}\right),
\]
\begin{equation}\label{plyatsko:60}
u^4=\frac{NV^4}{\Delta}\left(1+\frac{S_0^2 M}{m^2 r^3}\right).
\end{equation}
Similarly to (35), we note
\begin{equation}\label{plyatsko:61}
\varepsilon=\frac{|S_0|}{\mu r},
\end{equation}
where  $\varepsilon\ll 1.$

 The explicit expressions for $N$ we obtain directly from the condition
$u_\lambda u^\lambda=1$ in the form
\[
N=\Delta\left[\left(1+\varepsilon^2\frac{M}{r}\right)^2-3V_3
V^3\varepsilon^2\frac{M}{r}\times\right.
\]
\begin{equation}\label{plyatsko:62}
\left.\times\left(2-\varepsilon^2\frac{M}{r}\right)\right]^{-1/2}.
\end{equation}
Inserting (62) into (60) we obtain the expression for the
components $V^\lambda$ through $u^\lambda$ ($V^2=u^2=0$):
\[
V^1=u^1R\left(1-2\varepsilon^2\frac{M}{r}\right),
\]
\[
V^3=u^3R\left(1+\varepsilon^2\frac{M}{r}\right),
\]
\begin{equation}\label{plyatsko:63}
V^4=u^4R\left(1-2\varepsilon^2\frac{M}{r}\right),
\end{equation}
where
\begin{equation}\label{plyatsko:64}
R=\left[\left(1-2\varepsilon^2\frac{M}{r}\right)^2-3(u^3)^2\varepsilon^2
Mr\left(2-\varepsilon^2\frac{M}{r}\right)\right]^{-1/2}.
\end{equation}

The main feature of relationships (63) and (64) is that for the high
tangential velocity of a spinning particle the values $V^1, V^3,
V^4$ become imaginary. Indeed, if
\begin{equation}\label{plyatsko:65}
|u^3|>\frac{1}{\varepsilon\sqrt{6Mr}},
\end{equation}
in (64) we have the square root of the negative value. (As writing
(65) we neglect the small terms of order $\varepsilon^2$; all
equations in this section before (65)  are strict in
$\varepsilon.$) Using the notation for the particle's tangential
velocity $u_\perp\equiv ru^3$ by (65) we write
\begin{equation}\label{plyatsko:66}
|u_\perp|>\frac{\sqrt{r}}{\varepsilon\sqrt{6M}}.
\end{equation}
If $r$ is not much greater than $M$, the velocity value of
the right-hand side of equation (66) corresponds to the particle's
highly relativistic Lorentz $\gamma$-factor of order
$1/\varepsilon$.

This fact that according to (51), (63)--(66)  the
expressions for the components of 4-momentum $P^\lambda$ become
imaginary (if $\mu$ in (50) is real) is an evidence that the Tulczyjew-Dixon condition cannot be used for the particle's velocity which is very close to the velocity of light. In this context it is clear the sense of a result of recent paper [44] concerning some solutions of the MP equations at the Tulczyjew-Dixon condition which allow acceleration of a spinning particle to the superluminal velocity in Schwarzschild's background. This result is connected with the situation when relationship (66) takes place, i.e., when the Tulczyjew-Dixon condition  cannot be applied. Under the Mathisson-Pirani condition the corresponding result is not allowed by the MP equations.

\section{Highly relativistic solutions \\ of MP equations \\ in Kerr's background}

Some solutions of the MP equations under the Mathisson-Pirani condition in Kerr's background, which describe highly relativistic motions of a spinning particle, posses the feature common with the case of the corresponding motions in Schwarzschild's background, both for $a\ll M$ and $a\approx M$ (here $a$ and $M$ are the rotation parameter and the mass of the Kerr metric source; we put $a>0$). For example, far from the horizon surface, when $r\gg r_g$, the highly relativistic equatorial circular orbits of a spinning particle in Kerr's background are allowed practically with the same dependence of the $\gamma$-factor on $r$ as in Fig. 4
for the Schwarzschild background [23] (in [23] the Boyer-Lindquist coordinates are used). In this context we note that for growing values $r$ the spin-orbit and spin-spin interactions decrease as $1/r^3$ and $1/r^4$ correspondingly; therefore, for $r\gg r_g$  the influence of the spin-orbit action on a particle is prevailing . Other situations arise when a particle is not far from $r_g$. For example, Fig. 8 illustrates a case of the circular orbits in the narrow space domain near $r_{ph}^{(+)}$ and $r_{ph}^{(-)}$ (these values are the radial coordinates of the co- and counter-rotation photon circular orbits correspondingly) at the small value of the parameter $a$. In contrast to Fig. 3 for $a=0$, in Fig. 8 there is the upper graph which arise just due to $a\ne 0$, i.e., when spin-spin action is nonzero. Many other figures for different correlation of the signs $a$, $d\varphi/ds$, and $S_\theta$ are presented in [23].
\begin{figure}
\centering
\includegraphics[width=5cm]{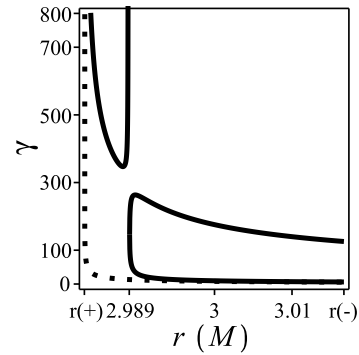}
\caption{\label{plyatsko_fig:8} Dependence of the Lorentz factor on $r$ for the highly relativistic equatorial circular orbits with $d\varphi/ds>0$ of the spinning particle in Kerr's background at $a=0.0145M$, $\varepsilon_0=10^{-2}$ , and $S_\theta>0$ (solid lines). The dotted line corresponds to the geodesic circular orbits. Here  we use the notation $r(-)$ for  $r_{ph}^{(-)}$ and   $r(+)$ for  $r_{ph}^{(+)}$.}
\end{figure}

Similarly to the case of Schwarzschild's background, the MP equations have the highly relativistic nonequatorial circular orbits in Kerr's background as well [14]. The space domain of these orbits existence increase with the increasing parameter $a$. If $a=M$, this domain is determined by $1.89M<r<4M$. The maximum value of the hovering angle
$\pi/2-\theta$ is achieved at $a=M$ and is equal to $\approx 63^{\circ}$, in contrast to $\approx 47^{\circ}$ at $a=0$.

Some graphs which describe nonequatorial motions (in particular, the $\theta$-oscillations) in Kerr's background when its spin is not orthogonal to the equatorial plane are presented in  [23].

\section{Numerical estimates}

Do some particles in cosmic rays posses the sufficiently high $\gamma$-factor for motions on the highly relativistic circular orbits, or on some their fragments, in the gravitational field of the Schwarzschild or Kerr black hole which are considered above? Yes, they do.
By the numerical estimates  for an electron in the gravitational field of a black hole with three of the Sun's mass the value $|\varepsilon_0$ is equal $4\times 10^{-17}$. Then the necessary value of the $\gamma$-factor for the realization of some highly relativistic circular orbits by the electron near this black hole is of order $10^8$. This $\gamma$-factor corresponds to the energy of the electron free motion of order $10^{14}$ eV. Analogously, for a proton in the field of such a black hole the corresponding energy is of order  $10^{18}$ eV. For the massive black hole those values are greater: for example, if $M$ is equal to $10^6$ of the Sun's mass the corresponding value of the energy for an electron is of order $10^{17}$ eV and for a proton it is $10^{21}$ eV.
Naturally, far from the black hole these values are greater because the necessary $\gamma$-factor is proportional to $\sqrt{r}$.

Note that for a neutrino near the black hole with three of the Sun's mass the necessary values of its $\gamma$-factor for motions on the highly relativistic circular orbits correspond to the neutrino's energy of the free motion of order $10^5$ eV. If the black hole's mass is of order $10^6$ of the Sun's mass, the corresponding value is of order $10^8$ eV.

Can the highly relativistic spin-gravity effects be registered by the observation of the electromagnetic synchrotron radiation from some black holes? Perhaps, however, it is difficult to differ the situation when the circular orbits of a spinning charged particle and its synchrotron radiation are caused by the magnetic field or the spin-gravity coupling. The detailed analysis of the observational data is necessary.

\section{On correspondence between \\ the Dirac and MP equations}

Here we point out a question which arise after the comparison of the MP equations with the general relativistic Dirac equation. As we note in the Introduction, the first equations follow from the second equation in some classical approximation. However, this result has been obtained in the linear spin approximation only, when expression (3) for the Fermi transported spin practically coincides with the corresponding expression for the parallel transport. Note that the general relativistic Dirac equation was obtained for eight years before the MP equations and therefore the Fermi transport was not discussed in the context of possible description of a quantum spinning particle in the gravitational field. At the same time, if one want to satisfy the principle of correspondence between an equation for a quantum spinning particle and the MP equations, now it is
necessary to take into account the known fact that according to the MP equations the spin of a test particle is transported by Fermi. However, in common sense, the notation "Fermi-transported spinor" cannot be introduced without violation of the Lorentz invariance. In this context it is interesting that for last years the possibility of the Lorentz invariance violation is discussed in the literature from different points of view. In any case (with violation of the  Lorentz invariance or not), it is necessary to propose a more exact equations for fermions in gravitational field than the usual general relativistic Dirac equation (probably, this corrected equation must be nonlinear in the $\psi$-function [45].

\section{Conclusions}

1. The MP equations are an important source (in the certain sense, even  the unique source) of information on  highly relativistic motions of spinning
particles in the gravitational field.

\noindent
2. Different solutions of the MP equations in Schwarzschild's and Kerr's background show clearly that the highly relativistic spin-gravity coupling can change significantly the trajectory of a
spinning particle as compare to the corresponding geodesic trajectory of a  spinless particle. In particular,  the highly relativistic spin-orbit interaction reveals the strong repulsive action on a spinning particle.

\noindent
3. In the highly relativistic regime of the spinning particle motions in the gravitational field the adequate supplementary condition for the MP equations is just the Mathisson-Pirani condition.

\noindent
4. By the numerical estimates, some particles in cosmic rays posses a sufficiently high $\gamma$-factor for motions on the significantly nongeodesic orbits near Schwarzschild's or Kerr's black holes.

\noindent
5. The MP equations can be useful  for searching adequate corrections in the known general relativistic Dirac equation with the aim to investigate the highly relativistic spin-gravity coupling on the quantum level.

\end{document}